\documentclass{pasj00}
\begin{document}
\SetRunningHead{Y. Takeda et al.}{Li, Na, and K Abundances in Sharp-Lined A-Type stars}
\Received{2011/09/29}%{yyyy/mm/dd}
\Accepted{2011/10/22}%{yyyy/mm/dd}

\title{Lithium, Sodium, and Potassium Abundances \\
in Sharp-Lined A-Type Stars
\thanks{Based on data collected at Bohyunsan Optical Astronomy Observatory
(KASI, Korea) and Okayama Astrophysical Observatory (NAOJ, Japan).}
%\thanks{The electronic table (table E) will be made available on-line
%at the PASJ web site upon publication, while it is provisionally placed at 
%$\langle$http://optik2.mtk.nao.ac.jp/\~{ }takeda/Aalkali/$\rangle$.}
}

%%% Please use the following style in case that sorting by 
%%% affilation is impossible. 
%
% \author{%
%   D-Firstname \textsc{D-Familyname}\altaffilmark{1}
%   E-Firstname \textsc{E-Familyname}\altaffilmark{1,2}
%   and
%   F-Firstname \textsc{F-Familyname}\altaffilmark{2}}
% \altaffiltext{1}{Address of Institute}
% \email{ddddd@xxx.xxx.xx.xx}
% \email{eeeee@xxx.xxx.xx.xx}
% \altaffiltext{2}{Address of Institute}

\author{
Yoichi \textsc{Takeda,}\altaffilmark{1}
Dong-Il \textsc{Kang,}\altaffilmark{2}
Inwoo \textsc{Han,}\altaffilmark{3}
Byeong-Cheol \textsc{Lee,}\altaffilmark{3}\\
Kang-Min \textsc{Kim,}\altaffilmark{3}
Satoshi \textsc{Kawanomoto,}\altaffilmark{1} and
Naoko \textsc{Ohishi}\altaffilmark{1}
}

\altaffiltext{1}{National Astronomical Observatory, 2-21-1 Osawa, 
Mitaka, Tokyo 181-8588}
\email{takeda.yoichi@nao.ac.jp, kawanomoto.satoshi@nao.ac.jp, naoko.ohishi@nao.ac.jp}
\altaffiltext{2}{Gyeongsangnamdo Institute of Science Education,\\
75-18 Gajinri, Jinsungmyeon, Jinju, Gyeongnam 660-851, Korea}
\email{kangdongil@gmail.com}
\altaffiltext{3}{Korea Astronomy and Space Science Institute,
61-1 Whaam-dong, Youseong-gu, Taejon 305-348, Korea}
\email{iwhan@kasi.re.kr, bclee@kasi.re.kr, kmkim@kasi.re.kr}

%%% Please use the following style in case that sorting by 
%%% affilation is impossible. 
%
% \author{%
%   D-Firstname \textsc{D-Familyname}\altaffilmark{1}
%   E-Firstname \textsc{E-Familyname}\altaffilmark{1,2}
%   and
%   F-Firstname \textsc{F-Familyname}\altaffilmark{2}}
% \altaffiltext{1}{Address of Institute}
% \email{ddddd@xxx.xxx.xx.xx}
% \email{eeeee@xxx.xxx.xx.xx}
% \altaffiltext{2}{Address of Institute}

%% `\KeyWords{}' always has to be placed before `\maketitle'.
%\KeyWords{xxxx:xxxx ......} %Do NOT move this preamble from here!

\maketitle

\begin{abstract}
The abundances of alkali elements (Li, Na, and K) 
were determined from the Li~{\sc i} 6708, Na~{\sc i} 5682/5688, 
and K~{\sc i}~7699 lines by taking into account the non-LTE effect
for 24 sharp-lined A-type stars ($v_{\rm e}\sin i \ltsim 50$~km~s$^{-1}$, 
7000~K~$\ltsim T_{\rm eff}\ltsim$~10000~K, many showing Am peculiarities 
to different degrees), based on high-dispersion and high-S/N spectral 
data secured at BOAO (Korea) and OAO (Japan). 
We found a significant trend that $A$(Na) tightly scales with $A$(Fe) 
irrespective of $T_{\rm eff}$, which means that Na becomes enriched 
similarly to Fe in accordance with the degree of Am peculiarity.
Regarding lithium, $A$(Li) mostly ranges between $\sim 3$ and $\sim 3.5$
(i.e., almost the same as or slightly less than the solar system 
abundance of 3.3) with a weak decreasing tendency with a lowering of 
$T_{\rm eff}$ at $T_{\rm eff} \ltsim 8000$~K, though several stars 
exceptionally show distinctly larger depletion.
The abundances of potassium also revealed an apparent $T_{\rm eff}$-dependence 
in the sense that $A$(K) in late-A stars tends to be mildly subsolar
(possibly with a weak anti-correlation with $A$(Fe))
systematically decreasing from ~$\sim 5.0$ ($T_{\rm eff} \sim 8500$~K) to 
$\sim 4.6$ ($T_{\rm eff} \sim 7500$~K), while those for early-A stars 
remain near-solar around $\sim$~5.0--5.2. 
These observational facts may serve as important constraints for any 
theory aiming to explain chemical anomalies of A-type stars.
\end{abstract}

%\section{}
%
%\noindent IMPORTANT NOTICE\\
%1. ``\verb|\draft|'' creates single column and double spaces format.\\
%2. If you comment out ``\verb|\draft|'', the output will be double column
%   and single space.\\
%3. For cross-references, the use of ``\verb|\label|, \verb|\ref|, \verb|\cite|'%' 
%   and the thebibliography environment is strongly recommended. \\
%4. Do NOT use ``\verb|\def|, \verb|\renewcommand|''.\\
%5. Do NOT redifine commands provided by PASJ00.cls.\\

%\newpage

%Sect. 1
\section{Introduction}

While elemental abundances in the photosphere of A-type stars,
which are of particular interest because of the chemical anomalies
of various types, have been extensively studied by a number 
of investigators so far, those of the alkali elements (Li, Na, 
and K) are barely known even nowadays, despite of their 
importance in stellar spectroscopy.

This is presumably due to the fact that these elements 
(with one valence electron weakly bound) are characterized by 
fairly low ionization potential ($\sim$~4--5~eV)
and most of them are ionized in the atmospheric condition 
of early-type stars, leaving only a tiny fraction as neutral.
Since detecting spectral lines of the dominant ionized species 
is almost hopeless (because of the closed-shell configuration 
requiring vary large excitation energy), one is obliged to 
measure the very weak lines of neutral species (Li~{\sc i}, Na~{\sc i}, 
K~{\sc i}). Thus, the necessity of using sufficiently high-S/N 
spectra must have hampered past researchers from investigating
the abundances of these elements.

Still, sodium has been comparatively well studied 
(e.g., Lane \& Lester 1987; Varenne \& Monier 1999; Gebran et al. 
2008, 2010; Gebran \& Monier 2008; Fossati et al. 2007, 2008)
mainly by using subordinate lines (Na~{\sc i} 5682/5688 or 6154/6160) 
of moderate strengths (i.e., neither too strong nor too weak), 
which are less affected by the non-LTE effect.  
However, Na abundance determinations in A-type stars conducted 
so far are restricted to sharp-lined late-A stars 
($T_{\rm eff} \ltsim 8500$~K) as well as F-type stars,
and thus Na compositions of early-A through late-B stars 
($T_{\rm eff} \gtsim 8500$~K) 
remain essentially unexplored, except for our previous non-LTE 
investigations for the very bright stars of Sirius 
(Takeda \& Takada-Hidai 1994) and Vega (Takeda 2008; hereinafter 
refereed to as Paper I), and the recent LTE studies of Fossati et al. 
(2009) for two sharp-lined late-B  stars (21~Peg and $\pi$~Cet).
Recently, Takeda et al. (2009; hereinafter referred to as Paper II)
carried out an extensive non-LTE analysis of Na~{\sc i} resonance
D lines at 5890/5896~$\rm\AA$ for a large (122) sample of
A-type stars (7000~K~$\ltsim T_{\rm eff} \ltsim 10000$~K)
including rapid as well as slow rotators, with an aim to
establish their Na abundance behaviors in general.
Unfortunately, however, since the D lines are so strong and the 
resulting abundances are quite sensitive to the microturbulent 
velocity $\xi$ , it turned out that ambiguities in the choice 
of $\xi$ (which is likely to be depth-dependent and difficult 
to assign a value relevant for the high-forming strong D lines) 
prevented from precise sodium abundance determinations. 

Almost the same situation holds for lithium: Determinations 
of Li abundances in A-type stars have been essentially restricted 
to late-A stars with $T_{\rm eff} \ltsim 8500$~K (e.g., Burkhart \& Coupry 
1991a, 1991b, 1995, 2000; Burkhart et al. 2005; North et al. 2005),
while early-A stars ($T_{\rm eff} \gtsim 8500$~K ) remain
practically untouched. Admittedly, this is a very difficult
and challenging task, as the Li~{\sc i}~6708 line considerably 
weakens as $T_{\rm eff}$ becomes higher. The work of Paper I
may be counted as an exceptional case, where the Li abundance of 
Vega could somehow be evaluated from the very weak Li~{\sc i}~6708 line 
($EW \ltsim 1$~m$\rm\AA$) based on a spectrum of ultra-high S/N ratio. 
Coupry and Burkhart (1992) reported the $EW$(Li~{\sc i}~6708) of 
$o$~Peg (a well-known very sharp-lined early-A star) to be 1.3~m$\rm\AA$,
which was used in Paper I (cf. Appendix~A therein) to argue that 
the Li abundance of $o$~Peg is almost consistent with the solar-system 
composition. However, as will be shown in this paper (cf. subsection 4.2), 
we consider that their measurement is an erroneous overestimation 
and Li is definitely underabundant in $o$~Peg. This example well 
illustrates the enormous difficulty of Li abundance derivation 
in hot A-type stars.

Investigations of potassium abundances are even more lacking.
As for late-A stars ($T_{\rm eff} \ltsim 8500$~K), it is (to our 
knowledge) only the study of Fossati et al. (2007) that 
reported the K abundances of late-A and Am stars in the Praesepe 
cluster, though their neglect of the non-LTE effect (which may be
significant for such a resonance line as K~{\sc i}~7699)
prevented from obtaining reliable results (cf. subsection 4.4
therein). Meanwhile, almost nothing has been reported
so far regarding the potassium abundances in early-A stars, except for 
Paper I where Vega's K abundance could be successfully established.

It is thus evident that our current knowledge on the behaviors 
of alkali-element abundances in A-type stars is considerably 
insufficient. Given this situation, we decided to carry out
a systematic spectroscopic study aiming at establishing the 
photospheric abundances of Na, Li, and K for an homogeneous
sample of A-dwarfs (7000~K~$\ltsim T_{\rm eff} \ltsim 10000$~K)
based on the spectral data of high quality.
Since measurements of very weak lines are concerned in the present 
case (especially for early-A stars), we had to confine ourselves
only to sharp-lined stars ($v_{\rm e}\sin i \ltsim 50$~km~s$^{-1}$).
This inevitably leads to the inclusion of a number of Am stars
(metallic-lined A stars), which are characterized by 
deficiencies in comparatively light elements (C, N, O, Ca, Sc)
as well as excesses of heavier elements (such as Fe-group or 
neutron-capture elements), since Am peculiarities are preferably 
observed in slow rotators (such as frequently seen in binaries).
Accordingly, the purpose of this paper is to clarify and discuss 
the behaviors of Na, Li, and K abundances in slowly-rotating A-type
stars based on the results of our analysis, while paying attention
to the possible connection (if any exists) between these alkali 
elements and the appearance/degree of Am anomalies.

%Sect. 2 

\section{Observational Data and Fundamental Parameters}

%subsect. 2.1 (table 1)
\subsection{Targets and Their Spectra}

We selected 21 sharp-lined stars\footnote{
We discarded HD~112185 ($\epsilon$~UMa; magnetic Ap star) and 
HD~40932 ($\mu$~Ori; double-lined binary) because of their
apparently unusual spectra, despite that these two stars have 
$v_{\rm e}\sin i$ values less than 50~km~s$^{-1}$.}
satisfying $v_{\rm e}\sin i \ltsim 50$~km~s$^{-1}$ from 122 A-type stars, 
studied in Paper II, where Na~{\sc i} 5890/5896 D lines were analyzed 
based on the spectra ($R\sim 45000$ and 
S/N ratios of a few hundreds) obtained by using BOES (Bohyunsan Observatory
Echelle Spectrograph) attached to the 1.8~m reflector at 
Bohyunsan Optical Astronomy Observatory (BOAO). We call these 21 stars
as ``BOAO sample.'' See section 2 of Paper II (and electronic table E
therein) for more details about these observational data.

Besides, in order to augment the spectral data for early-A stars
for which very high-quality is required to measure very weak lines
of neutral alkalis, we also used the spectra of 7 bright 
sharp-lined stars ($\pi$~Dra, $\alpha$~Lyr, $\alpha$~CMa, $\gamma$~Gem, 
15~Vul, 68~Tau, and $o$~Peg), which were observed with the HIDES
spectrograph (HIgh Dispersion Echelle Spectrograph; Izumiura 1999) 
attached to the 1.88~m reflector at Okayama Astrophysical 
Observatory (OAO). 
While the data of $\alpha$~Lyr were taken from the digital atlas
(based on the observations carried out on 2006 May 1--4)
published by Takeda, Kawanomoto, and Ohishi (2007) as in Paper I, 
those for the remaining 6 stars were secured by the observations
on 2008 October 4 (15~Vul, 68~Tau, $o$~Peg), October 7 ($\alpha$~CMa,
$\gamma$~Gem), and October 8 ($\pi$~Dra). These 2008 OAO data, which cover
the wavelength range of 4100--7800~$\rm\AA$ with three
mosaicked CCDs of 2K$\times$4K pixels, have high spectral
resolution ($R\sim 100000$) and very high S/N ratio (typically 
$\sim 1000$, or even higher for the very bright $\alpha$~CMa and $\gamma$~Gem).
We call these 7 stars as ``OAO sample.'' Note that the following 4 
stars are common to both of the BOAO and OAO samples:
HD~172167=$\alpha$~Lyr, HD~48915=$\alpha$~CMa, HD~47105=$\gamma$~Gem,
and HD~27962=68~Tau. Accordingly, the actual (net) number of the targets
is 24 ($=21+7-4$). The list of these program stars is presented
in table~1.

%subsect. 2.2 (figure 1, table 1)
\subsection{Atmospheric Parameters}

Regarding the effective temperature ($T_{\rm eff}$) 
and the surface gravity ($\log g$) of each program star,
we simply adopted the values used in Paper II for 21 BOAO sample
stars (including 4 OAO samples in common).
As for the remaining 3 OAO stars, those for 15~Vul and $o$~Peg 
were determined photometrically from Str\"{o}mgren's $b-y$, $c_{1}$, 
and $m_{1}$ colors along with the $\beta$ index by using Napiwotzki 
et al.'s (1993) calibration (as in Paper II; cf. subsection 3.1 
therein), while the values derived by Adelman (1996) were adopted 
for $\pi$~Dra (since color data were not available for this star).
The final values of $T_{\rm eff}$ and $\log g$ are summarized in
table 1. As can be seen in figure 1, where the targets 
are plotted on the theoretical HR diagram, the mass values of
the program stars range between $\sim$1.5~$M_{\odot}$ and 
$\sim $3~$M_{\odot}$.
The model atmospheres for each of the stars were constructed
by two-dimensionally interpolating Kurucz's (1993) ATLAS9 
model grid in terms of $T_{\rm eff}$ and $\log g$, where
we exclusively applied the solar-metallicity models 
as in Takeda et al. (2008a) as well as Paper II.

%Figure 1
\begin{figure}
  \begin{center}
    \FigureFile(50mm,50mm){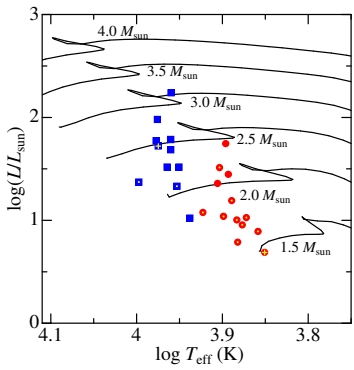}
    %%% \FigureFile(width,height){filename}
  \end{center}
\caption{
Plots of the program stars on the theoretical HR diagram
($\log (L/L_{\odot})$ vs. $\log T_{\rm eff}$), where the bolometric 
luminosity ($L$) was evaluated from the apparent visual magnitude with 
the help of the new Hipparcos parallax (van Leeuwen 2007) and 
Flower's (1996) bolometric correction. 
Circles (red) --- late A-type stars with $T_{\rm eff} < 8500$~K,
squares (blue) --- early A-type stars with $T_{\rm eff} > 8500$~K.
Yellow dots ($\cdot$) and crosses (+) are overplotted for 
classical Am stars (i.e., those classified as ``Am'' in the spectral
type given in the Bright Star Catalogue; cf. table 1) and Vega-like stars,
respectively, in order to discriminate them from others.
Theoretical evolutionary tracks corresponding to 
the solar metallicity computed by Girardi et al. (2000) for 
six different initial masses are also depicted for comparison.
}
\end{figure}

%subsect. 2.3 (figure 2, table 1)
\subsection{Rotational Velocity, Microturbulence, and Abundances of O/Fe}

Since we need to clarify the characteristics (e.g., parameters such as
rotation or atmospheric turbulence, degree of abundance peculiarities) 
of our sample stars, a synthetic spectrum fitting analysis 
based on Takeda's (1995) best-fit solution search algorithm was
applied to the 6146--6163~$\rm\AA$ region, in order to establish
the projected rotational velocity ($v_{\rm e}\sin i$),
microturbulent velocity dispersion ($\xi$), and the abundances of
iron and oxygen [$A$(Fe) and $A$(O); both are good indicators of
Am anomaly; cf. Takeda and Sadakane (1997)] (along with that of Na 
as a by-product), similarly to what was done in Takeda et al. (2008a; 
cf. section IV therein) and Paper II (cf. subsection 3.2 therein). 
Actually, we varied seven free parameters $\xi$, $v_{\rm e}\sin i$,  
$A$(O), $A$(Na), $A$(Si), $A$(Ca) and $A$(Fe) to accomplish the best fit,
where the non-LTE effect was taken into account for O (as in Takeda et al.
2010) as well as for Na (as in Paper II). The adopted atomic parameters of
important O, Na, and Fe lines relevant to this region are given in table 2.
The instrumental broadening corresponding to the spectral resolving 
power was also taken into account, since sharp-lined stars are concerned
here. 

Since we could not find a reasonable $\xi$ solution for the 
following five BOAO cases (i.e., stabilized at an apparently 
unreasonable value, or any convergence could not be accomplished),
we assumed an appropriate $\xi$ value as fixed and repeated the 
iterations until convergence.
$\xi= 2.5$~km~s$^{-1}$ (Adelman 1996; Monier 2005) for HD~95418,
$\xi= 2.1$~km~s$^{-1}$ (Caliskan \& Adelman 1997) for HD~43378,
$\xi= 3.0$~km~s$^{-1}$ (Sadakane 2006) for HD~218396,
$\xi= 3.9$~km~s$^{-1}$ (from the analytical formula derived by 
Takeda et al. 2008) for HD~204188, and
$\xi= 2.53$~km~s$^{-1}$ ($\xi$~solution from the OAO spectrum) for HD~48915.
The finally adopted results of $\xi$, $v_{\rm e}\sin i$, $A_{61}$(Fe) 
$A_{61}$(O), are presented in table 1 as well as in the electronic table E
(where $A_{61}$(Na) is also given).
How the theoretical and observed spectra match each other with 
the converged solutions of these parameters is shown in figure 2.

%Figure 2
\begin{figure}
  \begin{center}
    \FigureFile(50mm,200mm){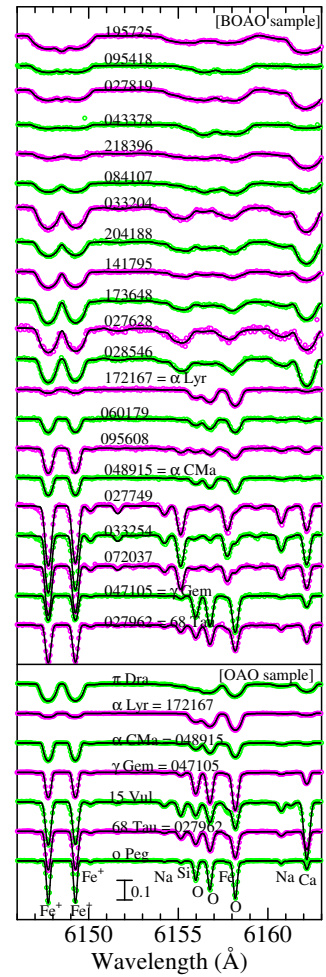}
    %%% \FigureFile(width,height){filename}
  \end{center}
\caption{
Synthetic spectrum fitting at the 6146--6163~$\rm\AA$ 
region while varying $\xi$, $v_{\rm e}\sin i$,  
$A$(O), $A$(Na), $A$(Si), $A$(Ca) and $A$(Fe).
The best-fit theoretical spectra are shown by solid lines, 
while the observed data are plotted by symbols.  
For each of the BOAO sample and OAO sample, the spectra are arranged 
(from top to bottom) in the descending order of $v_{\rm e}\sin i$ 
as in table 1, and an offset of 0.1 is applied to each spectrum 
relative to the adjacent one. 
}
\end{figure}

In figure 3, the resulting $A$(O) and $A$(Fe) values are plotted
against $v_{\rm e}\sin i$ and $T_{\rm eff}$, and their mutual 
correlation is also displayed.\footnote{
The reference standard abundances indicated in figures 3, 5, 8, 10, and 11
in this paper are the solar photospheric abundances (solar-system 
meteoritic abundance only for Li) taken from Grevesse and Noels 
(1993; for Li, O, Fe) and Anders and Grevesse (1989; for Na, K).}
We can see from these figures that the underabundance of O and 
the overabundance of Fe (both are the characteristics of 
Am peculiarities) are surely anti-correlated (figure 3c) and their weak 
$v_{\rm e}\sin i$-dependence is recognized (figures 3a,b) even in 
such a small range of $v_{\rm e}\sin i$  (i.e., larger anomaly for 
slower rotators; cf. Takeda \& Sadakane 1997).
We also notice that two stars (HD~218396 and HD~172167=Vega) are 
apparently metal-deficient ([Fe/H]$\sim -0.5$), showing the abundance 
charateristics of Vega-like stars (cf. Sadakane 2006), which are
presumably related to $\lambda$~Boo stars (see, e.g., Paunzen 2004
and the references therein).
It is interesting that these two appear to follow the O vs. Fe 
anti-correlation trend exhibited by Am stars, despite that
interaction with interstellar clouds (cf. Kamp \& Paunzen 2002;
Paunzen et al. 2002) is becoming a promising mechanism for
explaining the $\lambda$~Boo phenomenon (instead of the diffusion
process being considered as the standard theory for Am stars).
Regarding the microturbulence, the $\xi$ vs. $T_{\rm eff}$ relation 
depicted in figure 3e implies that the $\xi$ values resulting 
from the 6146--6163~$\rm\AA$ fitting are consistent with the 
analytical formula proposed by Takeda et al. (2008a) (which was 
invoked in Paper II).

%Figure 3
\begin{figure}
  \begin{center}
    \FigureFile(80mm,200mm){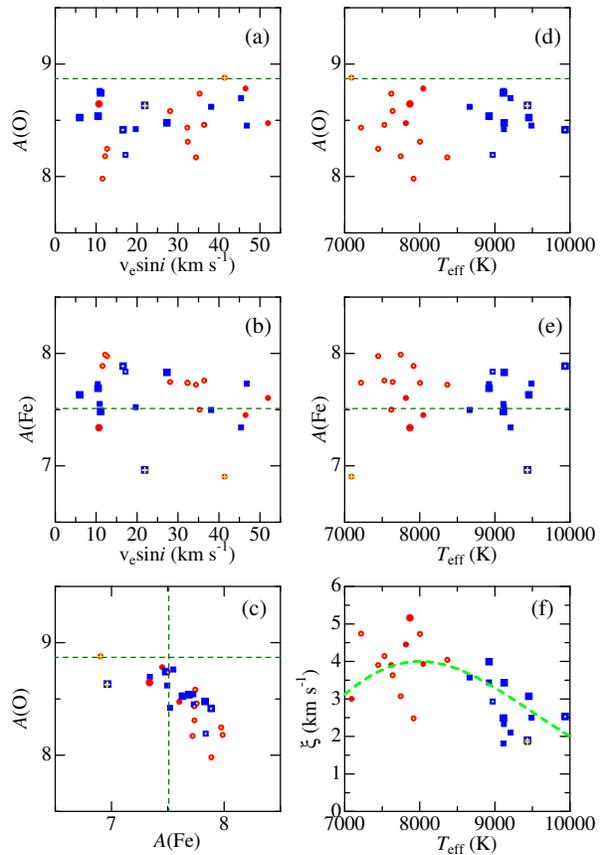}
    %%% \FigureFile(width,height){filename}
  \end{center}
\caption{
Characteristics and mutual correlations of the abundances/parameters
derived from the 6146--6163~$\rm\AA$ region fitting, given in table 1.
(a) $A$(O) (non-LTE oxygen abundance) vs. $v_{\rm e}\sin i$
(projected rotational velocity), (b) $A$(Fe) (LTE iron abundance)
vs. $v_{\rm e}\sin i$, (c) $A$(O) vs. $A$(Fe),
(d) $A$(O) vs. $T_{\rm eff}$, (e) $A$(Fe) vs. $T_{\rm eff}$,
and (f) $\xi$ (microturbulence) vs. $T_{\rm eff}$.
Circles (red) and squares (blue) correspond to stars with 
$T_{\rm eff} < 8500$~K, and those with $T_{\rm eff} > 8500$~K, 
respectively, where BOAO (smaller symbol) 
and OAO (larger symbol) results are discriminated by the symbol size.
As in figure 1, (yellow) dots and crosses are overplotted for
classical Am stars and Vega-like stars, respectively.
The reference solar abundances are indicated by
dashed lines in panels (a)--(e).
In panel (f), the analytical approximation derived in Takeda et al. 
(2008a), $\xi = 4.0 \exp\{- [\log (T_{\rm eff}/8000)/A]^{2}\}$~km~s$^{-1}$
(where $A \equiv [\log (10000/8000)]/\sqrt{\ln 2}$), 
is also depicted by the dashed curve for comparison.
}
\end{figure}

%subsect. 2.4 
\subsection{Parameter Uncertainties and Their Impact on Abundances}

In the next section (section 3) we will derive not only the abundances
(of Na, Li, and K) but also the relevant abundance errors in response to 
the possible ambiguities in the adopted atmospheric parameters, 
where we estimate $\pm 300$~K, $\pm 0.3$~dex, and $\pm 30$\% as 
the typical uncertainties in the absolute values of $T_{\rm eff}$, 
$\log g$, and $\xi$, respectively. While the internal errors of
our color-based $T_{\rm eff}$ and $\log g$ are $\sim 200$~K (2.5\%)
and $\sim 0.1$~dex according to Napiwotzki et al. (1993; cf. section 5
therein), we assumed somewhat larger ambiguities than these in view of 
the comparison with various published values (cf. section IV-c in Takeda
et al. 2008a). The uncertainty in $\xi$ was estimated from the 
dispersion\footnote{
This amount of dispersion ($\sim$~30\%) is regarded as a typical 
ambiguity in $\xi$, which can also be recognized by comparing various 
$\xi$ values published so far (see, e.g., figure 1 in 
Coupry \& Burkhart 1992; figure 2 in Gebran \& Monier 2007).
Yet, apart from this intrinsic dispersion, our $\xi$ values tend 
to be somewhat larger than the previously reported results (usually 
based on $EW$ values of Fe lines of various strengths). 
For example, our figure 3f suggests that the maximum $\xi$  
attained at mid-A type is $\sim$~4--5~km~s$^{-1}$, while it is
$\sim 3$~km~s$^{-1}$ in figure 2 of Coupry and Burkhart (1992);
also, our $\xi$ results are systematically higher by $\sim$~1~km~s$^{-1}$
with those of Landstreet et al. (2009), where 5 stars (68~Tau, $\gamma$~Gem, 
$\alpha$~CMa, 15~Vul, and $o$~Peg) are in common with our sample.
We suspect that this may be due to a depth-dependence of $\xi$
decreasing with height (cf. section 5 in Paper II). That is,
our $\xi$-determination (based on 6146--6163~$\rm\AA$ fitting) tends 
to reflect the physical condition in deeper layers (larger $\xi$) because 
of being mainly controled by the deep-forming O~{\sc i}~6155--8 triplet 
lines of high excitation, while the conventional method using 
$EW$s of high-forming strong lines may yield comparatively lower $\xi$.} 
in the $\xi$ vs. $T_{\rm eff}$ relation (figure 3f).

Since resonance lines of alkali elements with low-ionization potentials 
are concerned in the present case, 
it is comparatively easy to understand (at least qualitatively) 
how the abundance is affected by changing each parameter. 
That is, since almost all atoms are in the ground state of the 
first-ionized (closed-shell) stage, and only a small fraction of them 
remain neutral, the number population of the ground level of the 
neutral atoms ($n_{1}$, which is proportional to the line opacity $l$) 
is expressed as 
$n_{1} \propto \epsilon \theta^{3/2} n_{\rm e} 10^{\chi_{\rm I} \theta}$
according to Saha's equation ($\epsilon$ is the abundance, 
$\theta \equiv 5040/T$, $\chi_{\rm I}$ is the ionization potential
($\sim$~4--5~eV), and $n_{\rm e}$ is the electron density).
Then, the dependence of the line-opacity ($l$) upon $\theta_{\rm eff} 
(\equiv 5040/T_{\rm eff}$) and $g$ may be written as
$l \propto \epsilon \theta_{\rm eff}^{3/2} g^{\alpha} 10^{\chi_{\rm I} 
\theta_{\rm eff}}$
where we put $\theta \sim \theta_{\rm eff}$ and used the approximation 
$n_{\rm e} \propto g^{\alpha}$ ($\alpha \sim $~1/3--2/3 depending on 
$T_{\rm eff}$; see, e.g., Gray 2005).
Consequently, the abundance ($\epsilon$) resulting from a given
equivalent width is strongly dependent upon $\theta_{\rm eff}$ 
as $\epsilon \propto 
\theta_{\rm eff}^{-3/2}10^{-\chi_{\rm I}\theta_{\rm eff}}$
because of the exponential factor (i.e., $\epsilon$ increases 
as $T_{\rm eff}$ becomes higher).
The $\log g$-dependence is somewhat more complicated because the effect 
of the continuum opacity ($\kappa$) has to be taken into consideration.
For late A-type stars of comparatively lower $T_{\rm eff}$ 
($\sim$~7000--8000~K), H$^{-}$ opacity is still important and 
$\kappa$ has almost the same $g$-dependence as $l$; thus the abundances 
turn out rather independent on $g$ because of the cancellation in the 
$l/\kappa$ ratio determining the line strength (cf. Takeda et al. 2002).
Meanwhile, for early A stars of higher $T_{\rm eff}$ where 
Paschen continuum opacity of neutral hydrogen is dominant, 
$g$-dependence in $l$ directly reflects the line strength
and the abundance scales as $\epsilon \propto g^{-\alpha}$
(i.e., $\epsilon$ decreases as $\log g$ becomes higher).
Finally, regarding the effect of $\xi$ ($\epsilon$ tends to
decrease with an increase in $\xi$), the abundance results are 
practically inert to any change in $\xi$, since weak lines in the linear 
part of the curve of growth are concerned in most of the present cases 
(though only the Na abundances derived from Na~{\sc i} 5688 line 
for late A stars with $T_{\rm eff} \ltsim $~8000~K exceptionally show 
some appreciable $\xi$-dependence; cf. figure 5f).
We will show in section 3 that these expected trends can be actually 
confirmed by actual calculations (cf. figures 5, 8, and 10).

%Sect. 3 
\section{Abundance Determinations}

%subsect. 3.1 (figure 4)
\subsection{A(Na) from Na~I 5682/5688 fitting} 

The abundances of Na ($A_{61}$(Na)) were already obtained from the 
fitting analysis in the 6146--6163~$\rm\AA$ region (including Na~{\sc i} 
6154/6160 lines) as described in subsection 2.3, which are given 
in electronic table E. However, while the Na~{\sc i} 6154/6160
lines are quite suitable for Na abundance determination for late-A stars,
they become considerably weak as $T_{\rm eff}$ becomes higher and
the accuracy of $A_{61}$(Na) may be comparatively low at early-A stars, 
especially for the non-metal-rich case (actually, $A_{61}$(Na) of 
HD~172167=Vega in the BOAO sample could not be determined; cf. note 1 in 
electronic table E). 

We therefore decided to determine the Na abundances from the 
spectra in the 5680--5690~$\rm\AA$ region including
the Na~{\sc i} 5682/5688 lines (stronger than Na~{\sc i} 6154/6160
while not too strong to be seriously sensitive to $\xi$).
As done for the 6146--6163~$\rm\AA$ region, we applied the fitting 
procedure to the 5680--5690~$\rm\AA$ portion of the spectrum, 
where the telluric lines had been removed by dividing by 
the spectrum of a rapid rotator (cf. Paper I, Paper II).
We varied five abundance parameters of  
$A$(Na), $A$(Si), $A$(Sc), $A$(Fe) and $A$(Ni) to obtain
the best fit, while taking into account the non-LTE effect for Na
as in Paper II (cf. table 2 for the atomic parameters of Na~{\sc i} 
5682/5688). 
How the theoretical and observed spectra match each other with 
the converged solutions of these parameters is shown in figure 4.
The resulting sodium abundances, $A_{56}$(Na), are given in table 1.
Besides, as analysis-related quantities, the equivalent widths 
($EW$) inversely computed from the abundance, the corresponding 
non-LTE corrections ($\Delta$), and the abundance variations ($\delta$) 
in response to each of the perturbations in $T_{\rm eff}$ 
($\pm 300$~K), $\log g$ ($\pm 0.3$~dex), and $\xi$ ($\pm 30\%$), 
which were computed for the Na~{\sc i} 5688 line following the 
procedure described in Paper II (cf. subsections 4.3 and 4.4 therein), 
are plotted against $T_{\rm eff}$ in figure 5. 
The ($EW$, $\Delta$, $\delta$) results for both Na~{\sc i} 5682 and 
5688 lines are also presented along with $A_{56}$(Na) in 
electronic table E.

%Figure 4
\begin{figure}
  \begin{center}
    \FigureFile(50mm,200mm){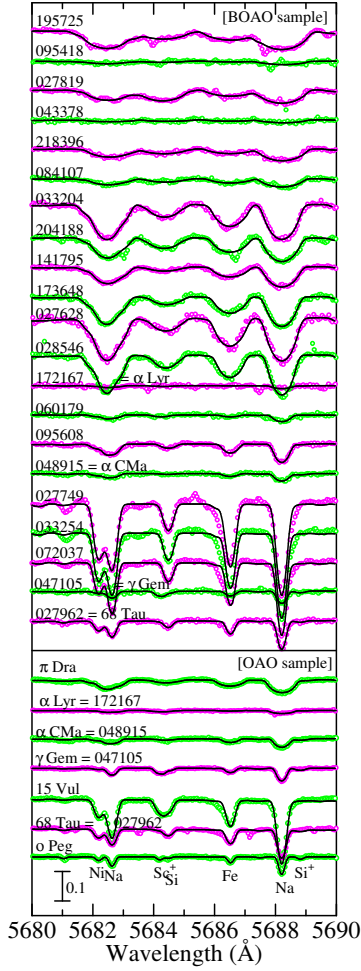}
    %%% \FigureFile(width,height){filename}
  \end{center}
\caption{
Synthetic spectrum fitting at the 5680--5690~$\rm\AA$ 
region while varying $v_{\rm e}\sin i$,  
$A$(Na), $A$(Si), $A$(Sc), $A$(Fe), and $A$(Ni).
The best-fit theoretical spectra are shown by solid lines, 
while the observed data are plotted by symbols.  
Otherwise, the same as in figure 2.
}
\end{figure}

%Figure 5
\begin{figure}
  \begin{center}
    \FigureFile(60mm,160mm){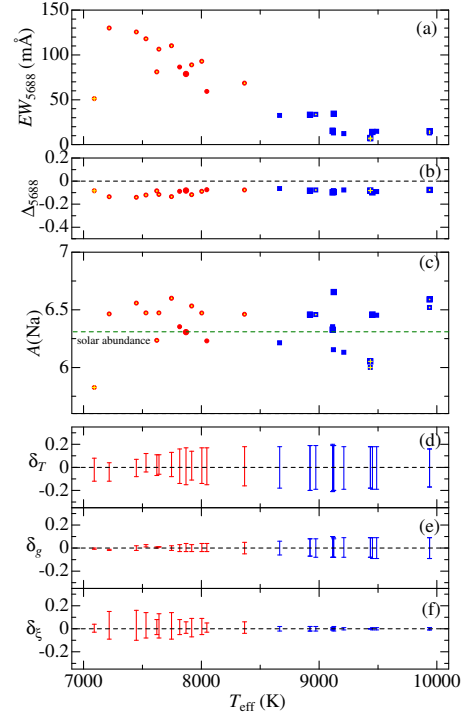}
    %%% \FigureFile(width,height){filename}
  \end{center}
\caption{
Sodium abundances derived from the synthetic spectrum fitting in the 
5680--5690~$\rm\AA$ region, along with the abundance-related quantities 
specific to the Na~{\sc i} 5688.21 line, plotted against $T_{\rm eff}$. 
(a) $EW_{5688}$ (equivalent width), 
(b) $\Delta_{5688}$ (non-LTE correction),
(c) $A$(Na) (non-LTE sodium abundance).
(d) $\delta_{T+}$ and $\delta_{T-}$ (abundance variations 
in response to $T_{\rm eff}$ changes of $+300$~K and $-300$~K), 
(e) $\delta_{g+}$ and $\delta_{g-}$ (abundance variations 
in response to $\log g$ changes of $+0.3$~dex and $-0.3$~dex), 
and (f) $\delta_{\xi +}$ and $\delta_{\xi -}$ (abundance 
variations in response to perturbing the standard $\xi$ value
by $\pm 30\%$). See subsection 2.4 for the description on 
the parameter uncertainties.
The signs of $\delta$'s are $\delta_{T+}>0$, $\delta_{T-}<0$,
$\delta_{g+}<0$, $\delta_{g-}>0$, $\delta_{\xi +}<0$, and
$\delta_{\xi -}>0$.
The meanings of the filled symbols are the same as in figure 3. 
}
\end{figure}

The comparison of the two sodium abundances [$A_{61}$(Na) and 
$A_{56}$(Na)] derived from different regions/lines is displayed
in figure 6. We can see from this figure that the agreement  
is good especially for late-A stars (red circles).
The star showing the largest departure ($\sim$~0.4 dex) is Vega 
(OAO sample) where $A_{61}$(Na)~=~5.63 and $A_{61}$(Na)~=~6.05 
were derived, which is due to the low accuracy of $A_{61}$ caused
by the extreme weakness of Na~{\sc i} 6154/6160 lines in this star 
(cf. figure 3 in Paper I).\footnote{
This result for Vega ($A_{61}$(Na)~=~5.63) based on the 
6146--6163~$\rm\AA$ region fitting is considerably
different from the $EW$-based value of 6.26 ($= 6.36-0.10$;
cf. table 1 in Paper I), despite that both are based essentially
on the same lines and the same non-LTE corrections. We understand that
the reason for this marked discrepancy is the erroneous underestimation of
the 6146--6163~$\rm\AA$ fitting-based result (5.63) in this study.
Namely, while the synthetic fitting method applied to a spectrum 
extending over some range (e.g., $\sim 10$~$\rm\AA$) is generally 
useful for simultaneous determinations of several elements at a time, 
it is not suitable for such a case where {\it extremely} weak line 
features are under question. That is, even a slight fluctuation of 
the continuum level can be a crucial source of error, because
an appropriate adjustment of the local continuum is impossible 
in such a global spectrum synthesis. Accordingly, when one has to
derive an abundance from an extremely weak line, the classical 
line-by-line $EW$ analysis (adopted in Paper I) should be preferable.
This is actually the reason why we adopted the classical $EW$-based
abundance determination for Li and K, because the lines of 
these elements become very weak in early-A stars.}
We hereafter adopt $A_{56}$(Na) as the representative sodium abundances 
which we will discuss in subsection 4.1.

%Figure 6
\begin{figure}
  \begin{center}
    \FigureFile(50mm,50mm){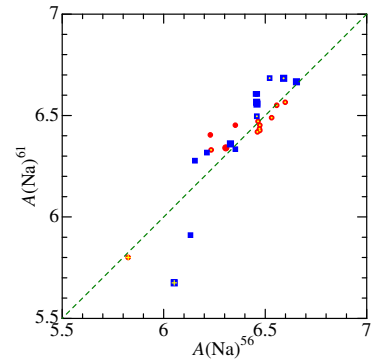}
    %%% \FigureFile(width,height){filename}
  \end{center}
\caption{
Comparison the (non-LTE) sodium abundances derived from the 6146--6163~$\rm\AA$ 
fitting [$A$(Na)$^{61}$] and those from the 5680--5690~$\rm\AA$ 
fitting [$A$(Na)$^{56}$; finally adopted in this study].
The same meanings of the symbols as in figure 3.
}
\end{figure}

%subsect. 3.2 (figure 7)
\subsection{A(Li) from Li~I 6708 equivalent width}

We derived the abundance of Li from the $EW$ of the merged
Li~{\sc i} doublet feature at $\sim 6708.8$~$\rm\AA$.
Since very weak lines are generally concerned, $EW$ 
measurements were carried out by the Gaussian fitting 
while comparing the spectrum on the computer screen with 
the theoretical spectrum synthesized with the known 
$v_{\rm e}\sin i$ as well as appropriately varied $A$(Li), by which 
we tried to minimize the possibility of erroneous measurement. 

In case we could not detect any suitable feature at the 
expected position, we gave up determination of $EW$.
Actually, we were unable to specify $EW$(6708) for about 
$\sim 50$\% of the targets (13 out of the 21 BOAO sample and 2 
out of the 7 OAO sample). In such cases, only the upper limit 
values were derived by the formula
\begin{equation}
EW^{\rm UL} \equiv k \times {\rm FWHM}/({\rm S/N})
\end{equation}
(cf. Takeda \& Kawanomoto 2005).
Here, FWHM was estimated from the root-sum-square of
(i) the rotational broadening
$w_{\rm r} (\simeq 2\times 0.78 \; v_{\rm e}\sin i \; \lambda /c)$
(cf. footnote 12 of Takeda et al. 2008b), (ii) the approximate separation 
of the doublet $d (\simeq 0.15\rm\AA)$, and (iii) the instrumental 
broadening $w_{\rm i} (\simeq R \lambda /c)$ ($R$: 
spectrum resolving power), as 
FWHM~$\equiv (w_{\rm r}^{2} + d^{2} + w_{\rm i}^{2})^{1/2}$,
and $k$ was assumed to be 2 according to our experience.
The actual measurements of $EW$(6708) were carried out on the 
``smoothed'' spectrum after applying the 9-pixel boxcar function 
to the original spectrum in order to reduce the effect 
of noise.\footnote{While how this smoothing process can improve
the appearance of the spectrum is demonstrated in figure 6 
of Paper I, the $EW$ value (0.7~m$\rm\AA$) of $\alpha$~Lyr 
at that time was actually measured in the ``original'' spectrum.
In this study, however, the measurement was done on the 
``smoothed'' spectrum, which is the reason for the appreciably 
larger value (1.2~m$\rm\AA$) derived for this star in spite 
of the use of the same spectrum data of Takeda et al. (2007).
This fact implies the difficulty in evaluating $EW$ of extremely 
weak line feature, which is subject to considerable uncertainties.}
Figure 7 displays how the Gaussian fitting for measuring $EW$ was 
done in the relevant Li~{\sc i}~6708 region of the spectra. 

%Figure 7
\begin{figure}
  \begin{center}
    \FigureFile(80mm,240mm){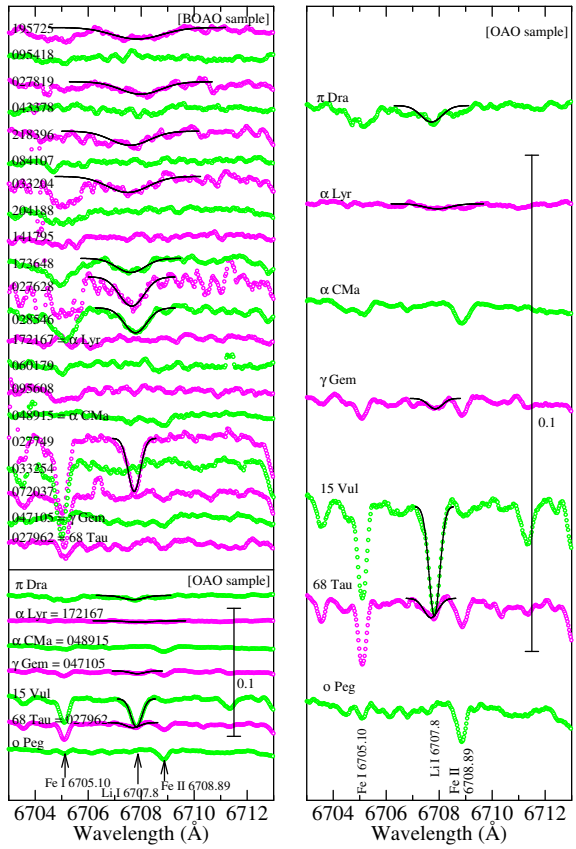}
    %%% \FigureFile(width,height){filename}
  \end{center}
\caption{
[Left panel] Observed spectra (symbols) in the 6703--6713~$\rm\AA$ region 
comprising the Li~{\sc i} 6708 line. For each of the BOAO and OAO samples, 
the spectra are arranged in the descending order of 
$v_{\rm e}\sin i$ and shifted by 0.02 relative to the adjacent one. 
The Gaussian profiles fitted for $EW$ measurements (when measurable) 
are depicted by solid lines. [Right panel] Spectra of seven OAO sample stars
as given in the lower region of the left panel, but with the magnified 
vertical scale for the purpose of recognizing very weak features.   
}
\end{figure}

Then, based on such evaluated $EW$ (or $EW^{\rm UL}$)
and the model atmosphere along with the non-LTE departure 
coefficients (appropriately interpolated from the precomputed 
ones on a model grid; cf. Takeda and Kawanomoto 2005 for more 
details of the non-LTE calculations), $A$(Li) (non-LTE abundance) 
or its upper limit was determined for each star (see table 2 for 
the adopted atomic parameters) as given in table 1.
Besides, in the similar manner as in subsection 3.1, we also 
computed the corresponding non-LTE corrections ($\Delta$), 
and the abundance variations ($\delta$) in response to 
perturbations of the atmospheric parameters. 
These [$EW$, $\Delta$, $A$(Li), $\delta$] results  
are plotted against $T_{\rm eff}$ in figure 8, 
which are also presented in electronic table E.

%Figure 8
\begin{figure}
  \begin{center}
    \FigureFile(50mm,160mm){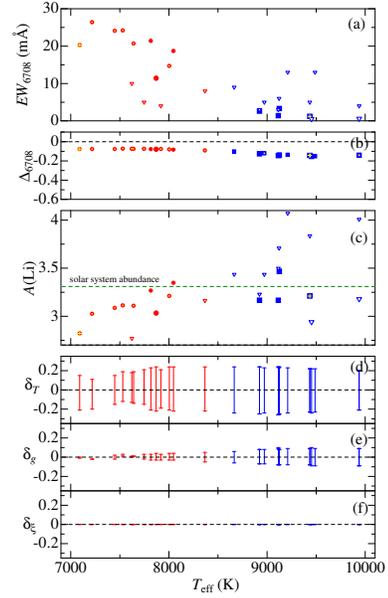}
    %%% \FigureFile(width,height){filename}
  \end{center}
\caption{
Lithium abundances, along with the abundance-related quantities 
specific to the Li~{\sc i} 6708 line (comprising several components), 
plotted against $T_{\rm eff}$. 
(a) $EW_{6708}$ (equivalent width), 
(b) $\Delta_{6708}$ (non-LTE correction),
(c) $A$(Li) (non-LTE lithium abundance),
The meanings of panels (d),(e), and (f) are the same as in figure 5. 
See the caption of figure 3 for the differences in type and size
of the filled symbols (circles and squares)
as well as for the meanings of overplotted dots and crosses. 
The open (inverse) triangles indicate the upper limit values 
for the non-detection cases. In panel (f), the extents of $\delta_{\xi}$ 
are so small (because of the extreme weakness of the line) that they 
are hardly recognizable to eyes.
}
\end{figure}

%subsect. 3.3
\subsection{A(K) from K~I 7699 equivalent width}

Regarding the abundance of K, we exclusively invoked the K~{\sc i} 
line at 7698.97~$\rm\AA$, since the alternative line of the doublet
at 7664.91~$\rm\AA$ tends to be severely influenced by strong
telluric lines. Its $EW$ was measured on the spectrum (in 
which telluric lines in the neighborhood had already been removed 
by dividing by that of a rapid rotator as in Paper I) in the same 
way as the case of the Li~{\sc i} 6708 line (cf. subsection 3.2). 
If the line failed to be detected (4 stars among the 21 BOAO sample), 
we estimated the upper limit ($EW^{\rm UL}$) following equation (1), 
where FWHM~$\equiv (w_{\rm r}^{2} + w_{\rm i}^{2})^{1/2}$ in this case.
Figure 9 displays how these measurements were done in the 
relevant K~{\sc i}~7699 region of the spectra.

%Figure 9
\begin{figure}
  \begin{center}
    \FigureFile(80mm,240mm){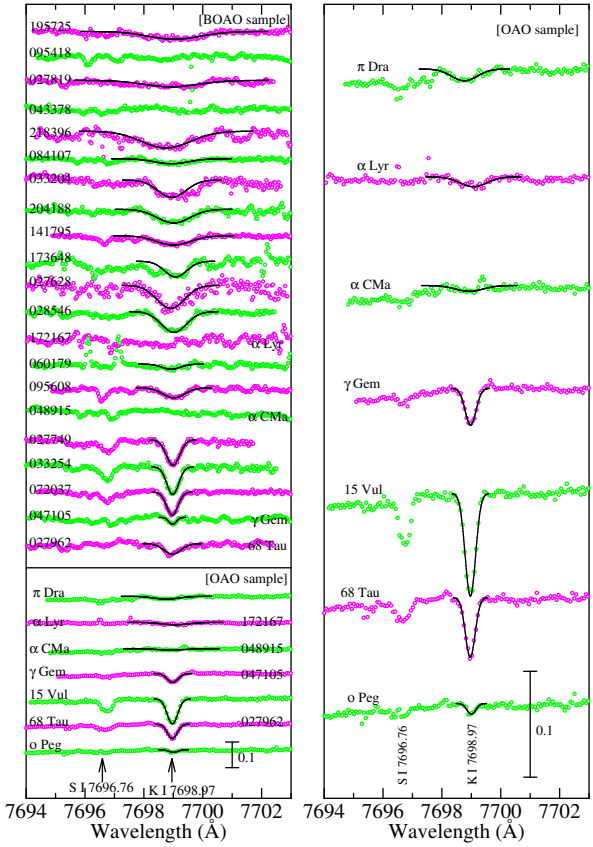}
    %%% \FigureFile(width,height){filename}
  \end{center}
\caption{
Observed spectra (symbols) in the 7694--7703~$\rm\AA$ region 
comprising the K~{\sc i} 7699 line. 
For each of the BOAO and OAO samples, the spectra are arranged in the descending 
order of $v_{\rm e}\sin i$ and shifted by 0.1 relative to the adjacent one. 
Otherwise, the same as in figure 7.
}
\end{figure}

Based on such evaluated $EW$ along with the non-LTE departure 
coefficients (appropriately interpolated from the precomputed 
ones on a model grid; cf. Takeda et al. 1996 for more 
details of the non-LTE calculations), the non-LTE K abundance ($A$(K))
or its upper limit was determined for each star (with the atomic 
parameter of the line in table 2), as given in table 1. These results 
of $EW$ and $A$(K), along with the corresponding non-LTE corrections 
($\Delta$) and the abundance variations ($\delta$) in response to 
perturbations of the atmospheric parameters, are plotted against 
$T_{\rm eff}$ in figure 10, and are also presented in electronic table E.

%Figure 10
\begin{figure}
  \begin{center}
    \FigureFile(50mm,160mm){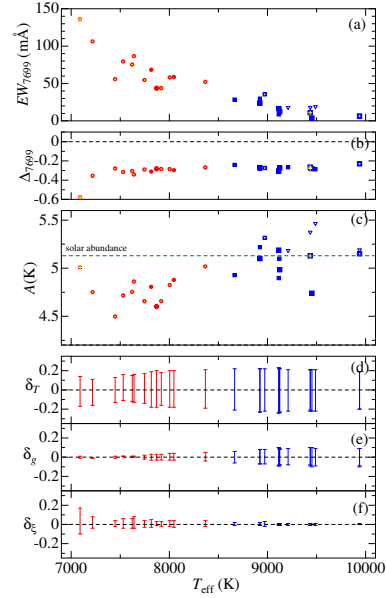}
    %%% \FigureFile(width,height){filename}
  \end{center}
\caption{
Potassium abundances, along with the abundance-related quantities 
specific to the K~{\sc i} 7698.97 line, plotted against $T_{\rm eff}$. 
(a) $EW_{7699}$ (equivalent width), 
(b) $\Delta_{7699}$ (non-LTE correction),
(c) $A$(K) (non-LTE potassium abundance).
The meanings of panels (d),(e), and (f), as well as the
discriminations of the symbols are the same as in figure 8. 
}
\end{figure}

%Sect. 4
\section{Discussion}

We are now ready to answer the questions which motivated this study 
(cf. section 1):
``How do the abundances of Na, Li, and K behave in sharp-lined A-type
stars? How are they related with the Am peculiarities 
frequently seen in slow rotators of this $T_{\rm eff}$ range?''
In figure 11 are plotted the resulting [$A$(Na), $A$(Li), and $A$(K)] 
against $A$(Fe) and $v_{\rm e}\sin i$ (determined in subsection 2.3), 
both of which are the key quantities closely connected with the Am phenomenon.
Based on this figure as well as figures 5c, 8c, and 10c (showing
the $T_{\rm eff}$-dependence), the abundance characteristics
of these elements are discussed in the following subsections, where 
we also compare the observed trend with the theoretical prediction from 
the atomic diffusion theory by Richer, Michaud, and Turcotte (2000)
as well as Talon, Richard, and Michaud (2006).

%Figure 11
\begin{figure}
  \begin{center}
    \FigureFile(80mm,200mm){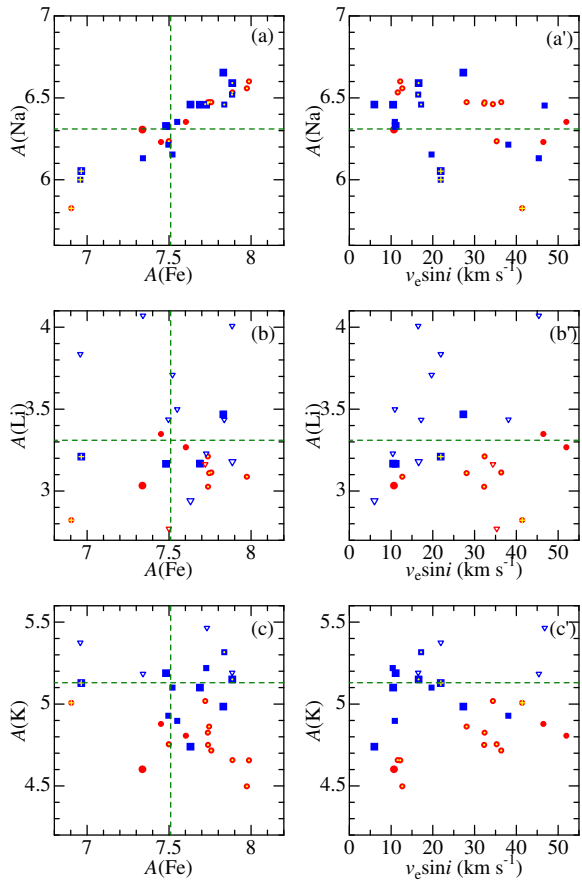}
    %%% \FigureFile(width,height){filename}
  \end{center}
\caption{
The abundances of Na, Li, and K plotted against the Fe abundance 
(left panels; a, b, c) and the projected rotational
velocity (right panels; a$'$, b$'$, c$'$).
The reference solar abundances (solar system abundance
for the case of Li) are indicated by dashed lines.
The meanings of the filled symbols 
(as well as the overplotted dots and crosses)
are the same as in figure 3, 
while the open (inverse) triangles indicate the upper limit values
for the non-detection cases.
}
\end{figure}

%subsect. 4.1
\subsection{Sodium}

Regarding Na, we notice in figure 11a a significant trend of 
remarkably tight correlation between $A$(Na) and $A$(Fe). 
Actually, a slight $v_{\rm e}\sin i$-dependence in $A$(Na) (similar to 
that seen for $A$(Fe) in figure 3b) is observed in figure 11a$'$.
Since $A$(Fe) is known to be one of the Am indicators  
in the sense that larger overabundance of Fe is accompanied by stronger 
Am stars, this observational fact means that Na behaves in accordance
with Fe (i.e., the excess of Na is a measure of Am anomaly just like
that of Fe).

While Na abundances of Am(+Fm) and normal A(+F) stars (though mostly 
restricted to those with $T_{\rm eff} \ltsim 8500$~K) in the field 
as well as in open clusters have been reported by several recent 
studies based on Na~{\sc i} 5683/5688 or 6154/6161 lines being less 
sensitive to non-LTE effects (e.g., Lane \& Lester 1987; Varenne \&
Monier 1999; Gebran et al. 2008, 2010; Gebran \& Monier 2008;
Fossati et al. 2007, 2008), nobody seems to have explicitly mentioned
this trend. However, an inspection of the results of these papers 
suggests tendencies of (i) moderately supersolar [Na/H] and (ii) 
[Na/H](Am)~$\gtsim$~[Na/H](normal A). Therefore, the existence of this 
positive correlation between [Na/H] and [Fe/H] appears almost certain.

From a theoretical point of view, however, such a conformance 
behavior between Na and Fe has never been predicted by any element 
segregation simulation designed for explaining Am peculiarities
such as that of Richer et al. (2000) or Talon et al. (2006), 
which suggests that [Na/H] remains almost solar or even slightly 
subsolar in spite of an appreciable overabundance of [Fe/H] 
(cf. figure 14 of Richer et al. or figure 16 of Talon et al.).
Therefore, the current diffusion theory (considered to be 
most promising for understanding the Am phenomena) is evidently 
imperfect at least in terms of Na, which remains to be improved. 
Here, the Na--Fe correlation revealed in this study would serve 
as an important observational constraint. Besides, it is interesting
that the apparently metal-deficient stars with $A$(Fe)~$\simeq$~6.9--7.0 
($\alpha$~Lyr and HD~218396=HR~8799) almost follow the same 
$A$(Na)--$A$(Fe) relation defined by Am and normal stars with
$7.3 \ltsim$~[Fe/H]~$\ltsim 8.0$. Since these two stars are
known to be Vega-like stars (which may be related to so-called
$\lambda$~Boo stars; see, e.g., Kamp \& Paunzen 2002, 
Paunzen et al. 2002), there might be some connection between
Am and $\lambda$~Boo anomalies. (See also the last paragraph
of subsection 2.3.)

%subsect. 4.2
\subsection{Lithium}

According to figures 11b and 11b$'$, we do not see any clear
$A$(Fe)- as well as $v_{\rm e}\sin i$-dependence in $A$(Li)
which has values around or slightly lower than the solar-system
abundance of $\sim 3.3$.
This may suggest that Am and normal A-type stars do not show
any manifest difference in terms of their surface Li abundances.

We recognize, however, a tendency of $T_{\rm eff}$-dependence
for late A-type stars in the sense that $A$(Li) gradually decreases 
from $\sim 3.3$ ($T_{\rm eff} \sim 8000$~K) to
$\sim 3.0$ ($T_{\rm eff} \sim 7000$~K)
while such a systematic trend is not seen for early A-type stars
($T_{\rm eff} \gtsim 8500$~K) where $A$(Li) tends to be around
the solar system abundance of $\sim 3.3$ (figure 8c). 
This $T_{\rm eff}$-dependent tendency of $A$(Li) in late-A stars 
is also observed  in figure 2 of North et al. (2005),
and is more or less consistent with the theoretical prediction
by Richer et al. (2000; cf. figure 14 therein).

It should be noted, however, that several stars apparently deviate 
from this general trend and show considerable Li-deficiencies (i.e., only 
the upper limit was derived since the Li line could not be detected).
Here, especially important are the following 4 stars: 
HD~204188 (7622~K, $<2.77$), HD~141795 (8367~K, $<3.16$),
$o$~Peg (9453~K, $<2.94$), and $\alpha$~CMa (9938~K, $<3.18$).
What makes such an exceptional depletion of Li?
All these are classified as Am stars ($o$~Peg is also an Am star
according to Adelman 1988, though the conventional type of A1~IV 
is given in table 1), though Am alone can not be the condition
to trigger such a marked underabundance of Li, because there are 
a number of other Am stars that do show normal behavior of Li.
Actually, most of the present sample of low $v_{\rm e}\sin i$ stars 
more or less show some degree of chemical peculiarities. 
In this connection, Burkhart and Coupry (1991a) stated (based on their 
study of late A-type stars) that ``occurrence of Li-deficient stars 
is not exceptional without any clearcut division between normal and Am stars.''
Yet, as far as early A-type stars are concerned, the fact that
Li tends to remain almost unaffected in near-normal stars such as $\gamma$~Gem 
(believed to have almost the solar abundances; e.g., Adelman \& Philip
1996, Lehmann et al. 2002) and $\pi$~Dra (which were previously
considered as normal even though later found to have only weak 
Am anomaly; cf. Sadakane \& Okyudo 1990) might suggest the existence 
of some connection between the conspicuous Li deficit of $\alpha$~CMa 
or $o$~Peg and their Am phenomenon.

Here, we have to remark that the conclusion of Paper I (the photospheric 
Li abundance of $o$~Peg almost coincides with that of the solar-system 
composition; cf. Appendix A therein), which was based on the 
$EW$(Li~6708) value of 1.3~m$\rm\AA$ taken from table 1 of 
Coupry and Burkhart (1992), was not correct.
As can be seen from figure 7, the Li~{\sc i}~6708 line is not 
detectable even on the spectrum of S/N~$\sim 1000$;
and we derived the upper limit of its equivalent width as
$EW$(6708)~$\ltsim 0.5$~m$\rm\AA$ (cf. the electronic table E) 
following equation (1).
We suspect, therefore, that Coupry and Burkhart (1992) 
misidentified/mismeasured this line, since the quality of their spectrum 
with typical S/N of $\sim 300$ (cf. section 1 therein) is evidently 
insufficient for measuring such an extremely weak line.
Our revised conclusion for $o$~Peg is $A$(Li)~$\ltsim 2.94$, 
which means that its photospheric Li abundance is at least by 
$\gtsim 0.4$~dex deficient compared to the solar-system value.

%subsect. 4.3
\subsection{Potassium}

Regarding $A$(K), we can observe (from figure 10c) a trend in terms 
of $T_{\rm eff}$ just similar to the case of $A$(Li): $A$(K) of 
late A-type stars appears to systematically decrease from $\sim 5.0$ 
($T_{\rm eff} \sim 8500$~K) to $\sim 4.6$ ($T_{\rm eff} \sim 7500$~K), 
while such a systematic trend is not seen for early A-type stars 
($T_{\rm eff} \gtsim 8500$~K) where $A$(K) is almost solar around 
$\sim 5.0$ (an exception is $o$~Peg which shows a clearly subsolar 
abundance of $\sim 4.7$). 
Meanwhile, it also appears that $A$(K)'s of late A-type stars are
weakly dependent upon $A$(Fe) as well as $v_{\rm e}\sin i$
in the sense that K gets more deficient as the Am anomaly becomes 
stronger (cf. the red circles in figures 11c and 11c$'$). 
Accordingly, the deficiency of K (particularly seen in late-A stars
of $T_{\rm eff} \ltsim 8500$~K) may be controlled by two factors 
($T_{\rm eff}$ and the degree of Am peculiarity).

Richer et al's (2000) simulation based on the theory of atomic diffusion 
(along with radiative acceleration) predicts a small overabundance
of K at higher $T_{\rm eff}$ ($\sim 10000$~K) region, whereas a
marginal underabundance is expected at the lower $T_{\rm eff}$ side 
($\sim 7000$~K) (cf. figure 14 therein). Such a predicted trend of 
positive $dA$(K)/$dT_{\rm eff}$ gradient is in qualitative agreement 
with what we found from figure 10c.

As mentioned in section 1, K-abundance determinations of A-type stars 
have so far been tried only by Fossati et al. (2007) in their LTE 
analysis of the K~{\sc i} 7699 line for late-A stars in the Praesepe 
cluster. According to their results for 5 stars, we see
$\langle$[K/H]$\rangle_{\rm LTE}\sim +0.2$~$(\pm 0.3)$
(cf. figure 10 and table 3 therein).  Considering that the typical 
(negative) non-LTE corrections amount to $\sim$~0.3--0.4~dex at the 
relevant $T_{\rm eff}$ range (cf. figure 10b), we may state
that their results would turn out to be mildly subsolar when the
non-LTE effect is taken into consideration, and become more consistent 
with our conclusion.

%Sect. 5
\section{Conclusion}

Motivated by the fact that our current understanding on the abundances 
of alkali elements in A-type stars is considerably insufficient,
we conducted a comprehensive non-LTE analysis
to establish the photospheric abundances of Na, Li, and K for 
24 selected sharp-lined A-type stars 
($v_{\rm e}\sin i \ltsim 50$~km~s$^{-1}$, 
7000~K~$\ltsim T_{\rm eff}\ltsim$~10000~K),
many of which are Am stars showing different degree of 
chemical peculiarity.

For this purpose, we primarily invoked the spectra of moderately 
high-dispersion ($R\sim 45000$) and high S/N ratio (typically a few 
hundreds) obtained with BOES at BOAO/Korea, though spectra of much higher 
quality ($R\sim 100000$ and S/N~$\gtsim 1000$) secured with HIDES
at OAO/Japan were additionally used for 7 apparently bright stars
(mostly of early A-type).

We first carried out spectrum fitting analyses applied to
the 6146--6163~$\rm\AA$ region and derived $\xi$, $v_{\rm e}\sin i$, 
$A$(O), and $A$(Fe), from which the degree of Am characteristics
for each star could be confirmed.

The abundances of sodium, which were determined by the spectrum fitting
method applied to the region comprising Na~{\sc i} 5682/5688 lines,
revealed a significant trend that $A$(Na) tightly scales with $A$(Fe), 
which means that Na becomes enriched 
similarly to Fe in accordance with the degree of Am phenomenon.
This result should be regarded as important because it seriously 
contradicts the prediction from the atomic diffusion theory.

Regarding lithium, $A$(Li) derived from the weak Li~{\sc i} line 
at 6708~$\rm\AA$ showed a tendency of $T_{\rm eff}$-dependence
for late A-type stars in the sense that $A$(Li) gradually decreases 
from $\sim 3.3$ ($T_{\rm eff} \sim 8000$~K) to
$\sim 3.0$ ($T_{\rm eff} \sim 7000$~K), while $A$(Li) for early A-type 
stars ($T_{\rm eff} \gtsim 8500$~K) tends to be around
the solar system abundance of $\sim 3.3$.
However, several stars apparently deviating from this general trend 
and showing marked underabundances do exist, though what triggers
such a Li deficiency is not clear.

The abundances of potassium derived from the K~{\sc i} 7699 line
revealed a tendency somewhat similar to the case of lithium:
An apparent $T_{\rm eff}$-dependence (and possibly with a weak 
anti-correlation with Am peculiarity) exists for late-A stars
where $A$(K) tend to be mildly subsolar systematically decreasing 
from ~$\sim 5.0$ ($T_{\rm eff} \sim 8500$~K) to $\sim 4.6$ 
($T_{\rm eff} \sim 7500$~K), whereas those for most early-A stars 
remain near-solar around $\sim$~5.0--5.2.

When these observational facts are compared with the theoretical 
prediction of Richer et al. (2000) or Talon et al. (2006) based on 
their atomic diffusion calculations, 
a serious discrepancy is found for Na (almost no anomaly 
or slightly subsolar tendency is predicted, while an enrichment of Na 
just like Fe is suggested from our analysis), though the 
simulated results appear to qualitatively reproduce the general 
trends observed for Li and K.
Further improvement and development on the theoretical side would be 
desirably awaited to explain the results concluded in this study,
which may serve as important constraints for any theory aiming 
to account for the origin of chemical anomalies in Am stars.

\bigskip

This research has made use of the SIMBAD database, operated by
CDS, Strasbourg, France. 
I. Han acknowledges the financial support for this study by KICOS through 
Korea--Ukraine joint research grant (grant  07-179).
B.-C. Lee acknowledges the Astrophysical Research Center for the Structure 
and Evolution of the Cosmos  (ARSEC, Sejong University) of the 
Korea Science and Engineering Foundation (KOSEF) through the Science
Research Center (SRC) program.

\clearpage

\onecolumn

%Table 1
%\clearpage
\setcounter{table}{0}
\begin{table}[h]
%\scriptsize
\small
\caption{Basic data of the program stars and the results of the analysis.}
\begin{center}
\begin{tabular}
{rrccccrrrrcccl}\hline \hline
HD\# & HR\# & Name & Sp.type & $T_{\rm eff}$ & $\log g$ & $\xi$ & $v_{\rm e}\sin i$ &
 $A_{\rm Fe}$ & $A_{\rm O}$ & $A_{\rm Li}$ & $A_{\rm Na}$ & $A_{\rm K}$ & Remark \\
\hline
\multicolumn{14}{c}{[BOAO sample]}\\
195725 & 7850& $\theta$ Cep & A7III  &  7816 & 3.74 & 4.45 & 52.0 & 7.60 & 8.47 &   3.27 & 6.35&  4.81  & \\ 
 95418 & 4295& $\beta$ UMa & A1V    &  9489 & 3.85 & $^{\dagger}$2.50 & 46.8 & 7.73 & 8.45 &  ($<$4.34) & 6.45& ($<$5.47) & \\ 
 27819 & 1380& $\delta^{2}$Tau & A7V    &  8047 & 3.95 & 3.93 & 46.5 & 7.45 & 8.78 &   3.35 & 6.23&  4.88 & H \\ 
 43378 & 2238& 2 Lyn   & A2Vs   &  9210 & 4.09 & $^{\dagger}$2.10 & 45.4 & 7.34 & 8.70 &  ($<$4.07) & 6.13& ($<$5.18) & \\ 
218396 & 8799&  $\cdots$  & A5V    &  7091 & 4.06 & $^{\dagger}$3.00 & 41.4 & 6.90 & 8.88 &   2.82 & 5.83&  5.01 & Vega-like star\\ 
 84107 & 3861& 15 Leo  & A2IV   &  8665 & 4.31 & 3.57 & 38.1 & 7.50 & 8.62 &  ($<$3.44) & 6.21&  4.93 & \\ 
 33204 & 1670&  $\cdots$ & A5m    &  7530 & 4.06 & 4.14 & 36.4 & 7.76 & 8.46 &   3.11 & 6.47&  4.72 & H \\ 
204188 & 8210&  $\cdots$ & A8m    &  7622 & 4.21 & $^{\dagger}$3.90 & 35.3 & 7.50 & 8.74 &  ($<$2.77) & 6.24&  4.75 & \\ 
141795 & 5892& $\epsilon$ Ser & A2Vm    &  8367 & 4.24 & 4.04 & 34.4 & 7.72 & 8.17 &  ($<$3.16) & 6.46&  5.02 & \\ 
173648 & 7056& $\zeta^{1}$Lyr & A4m     &  8004 & 3.90 & 4.73 & 32.4 & 7.74 & 8.31 &   3.21 & 6.47&  4.82 & \\ 
 27628 & 1368& 60 Tau  & A3m    &  7218 & 4.05 & 4.74 & 32.3 & 7.74 & 8.43 &   3.03 & 6.46&  4.75 & H \\ 
 28546 & 1428& 81 Tau  & A5m     &  7640 & 4.17 & 3.63 & 28.1 & 7.74 & 8.58 &   3.11 & 6.47&  4.86 & H \\ 
172167 & 7001& $\alpha$ Lyr & A0Va &  9435 & 3.99 & 1.85 & 21.9 & 6.96 & 8.63 &  ($<$3.84) & 6.00& ($<$5.38) & Vega-like star\\ 
 60179 & 2891& $\alpha$ Gem & A1V   &  9122 & 3.88 & 2.33 & 19.7 & 7.52 & 8.42 &  ($<$3.71) & 6.15&  5.10 & \\ 
 95608 & 4300& 60 Leo  & A1m    &  8972 & 4.20 & 2.93 & 17.2 & 7.84 & 8.19 &  ($<$3.43) & 6.46&  5.32 & \\ 
 48915 & 2491& $\alpha$ CMa & A1Vm &  9938 & 4.31 & $^{\dagger}$2.53 & 16.5 & 7.89 & 8.41 &  ($<$4.01) & 6.52& ($<$5.19) & \\ 
 27749 & 1376& 63 Tau  & A1m    &  7448 & 4.21 & 3.90 & 12.7 & 7.97 & 8.25 &   3.09 & 6.56&  4.50 & H \\ 
 33254 & 1672& 16 Ori  & A2m    &  7747 & 4.14 & 3.07 & 12.2 & 7.99 & 8.18 &  ($<$2.53) & 6.60&  4.66 & H \\ 
 72037 & 3354& 2 UMa   & A2m    &  7918 & 4.16 & 2.48 & 11.6 & 7.89 & 7.98 &  ($<$2.55) & 6.53&  4.66 & \\ 
 47105 & 2421& $\gamma$ Gem & A0IV   &  9115 & 3.49 & 1.81 & 10.9 & 7.55 & 8.76 &  ($<$3.50) & 6.35&  4.90 & \\ 
 27962 & 1389& 68 Tau  & A2IV   &  8923 & 3.94 & 3.44 & 10.4 & 7.73 & 8.54 &  ($<$3.23) & 6.46&  5.22 & H \\ 
\hline
\multicolumn{14}{c}{[OAO sample]}\\
182564 & 7371& $\pi$ Dra  & A2IIIs &  9125 & 3.80 & 3.43 & 27.3 & 7.83 & 8.48 &   3.47 & 6.65&  4.98 & \\ 
172167 & 7001& $\alpha$ Lyr & A0Va &  9435 & 3.99 & 1.90 & 21.9 & 6.97 & 8.63 &   3.21 & 6.05&  5.13 & Vega-like star\\ 
 48915 & 2491& $\alpha$ CMa & A1Vm  &  9938 & 4.31 & 2.53 & 16.6 & 7.89 & 8.42 &  ($<$3.18) & 6.59&  5.15 & \\ 
 47105 & 2421& $\gamma$ Gem & A0IV   &  9115 & 3.49 & 2.49 & 11.1 & 7.48 & 8.74 &   3.17 & 6.33&  5.19 & \\ 
189849 & 7653& 15 Vul  & A4III  &  7870 & 3.62 & 5.16 & 10.7 & 7.34 & 8.65 &   3.03 & 6.31&  4.60 & \\ 
 27962 & 1389& 68 Tau  & A2IV   &  8923 & 3.94 & 3.99 & 10.5 & 7.69 & 8.54 &   3.17 & 6.46&  5.10 & H \\ 
214994 & 8641& $o$ Peg & A1IV   &  9453 & 3.64 & 3.07 &  6.0 & 7.63 & 8.52 &  ($<$2.94) & 6.46&  4.74 & \\ 
\hline
\end{tabular}
\end{center}
%\scriptsize
Note. \\
In columns 1 through 6 are given the HD number, HR number, star name
(with constellation), spectral type (taken from Hoffleit 1982), effective 
temperature (in K), and logarithmic surface gravity (in cm~s$^{-2}$). 
Columns 7 through 10 show the results derived from 6146--6163~$\rm\AA$ region 
fitting: the microturbulent velocity (in km~s$^{-1}$), projected 
rotational velocity (in km~s$^{-1}$), (LTE) abundance of Fe, 
and (non-LTE) abundance of O. The non-LTE abundances of Li, Na, 
and K finally resulting from this investigation are presented in 
columns 11--13.  
All abundance results are expressed in the usual normalization 
of $A$(H)~=~12.00.
In each of the BOAO ans OAO samples, the objects are arranged 
in the descending order of $v_{\rm e} \sin i$. 
In the remark in column 14, Hyades cluster stars are denoted with ``H''.\\
$^{\dagger}$Regarding these $\xi$ values, literature values were adopted 
unlike the others, because the $\xi$-determination from the 
6146--6163~$\rm\AA$ region fitting did not work successfully. 
See subsection 2.3 for more details.
\end{table}

%Table 2
%\clearpage
\setcounter{table}{1}
\begin{table}[h]
%\scriptsize
\small
\caption{Atomic data of important lines relevat to the analysis.}
\begin{center}
\begin{tabular}
{cclrl}\hline \hline
Desig. & Species & $\lambda (\rm\AA) $ & $\chi_{\rm low}$~(eV)  & $\log gf$ \\
\hline
 Li~6708 & Li~{\sc i} &  6707.756 & 0.00 & $-0.43$ \\ 
         & Li~{\sc i} &  6707.768 & 0.00 & $-0.21$ \\
         & Li~{\sc i} &  6707.907 & 0.00 & $-0.93$ \\
         & Li~{\sc i} &  6707.908 & 0.00 & $-1.16$ \\
         & Li~{\sc i} &  6707.919 & 0.00 & $-0.71$ \\
         & Li~{\sc i} &  6707.920 & 0.00 & $-0.93$ \\
\hline
O~6156 & O~{\sc i} &  6155.961 & 10.74 & $-1.40$ \\
       & O~{\sc i} &  6155.971 & 10.74 & $-1.05$ \\
       & O~{\sc i} &  6155.989 & 10.74 & $-1.16$ \\
O~6157 & O~{\sc i} &  6156.737 & 10.74 & $-1.52$ \\
       & O~{\sc i} &  6156.755 & 10.74 & $-0.93$ \\
       & O~{\sc i} &  6156.778 & 10.74 & $-0.73$ \\
O~6158 & O~{\sc i} &  6158.149 & 10.74 & $-1.89$ \\
       & O~{\sc i} &  6158.172 & 10.74 & $-1.03$ \\
       & O~{\sc i} &  6158.187 & 10.74 & $-0.44$ \\
\hline
Na~5682& Na~{\sc i} & 5682.633  & 2.10 & $-0.70$ \\
Na~5688& Na~{\sc i} & $5688.20^{\dagger}$  & 2.10 & $-0.40^{\dagger}$ \\
Na~6154& Na~{\sc i} & 6154.226  & 2.10 & $-1.56$ \\
Na~6160& Na~{\sc i} & 6160.747  & 2.10 & $-1.26$ \\
\hline
Si~5684& Si~{\sc i} & 5684.484 & 4.95 & $-1.65$ \\
Si~5688& Si~{\sc ii} & 5688.817 & 14.19 & +0.40 \\
Si~6155& Si~{\sc i} & 6155.134 & 5.62 & $-0.40$ \\
\hline
S~7696 & S~{\sc i} & 7696.758 & 7.87 & $-0.89$ \\
\hline
K~7699 & K~{\sc i} &  7698.974 & 0.00  & $-0.17$ \\
\hline
Ca~6162 & Ca~{\sc i} & 6162.173 & 1.90 & +0.10 \\
\hline
Sc~5684 & Sc~{\sc ii} & 5684.202 & 1.51 & $-1.05$ \\
\hline
Fe~5686& Fe~{\sc i} & 5686.524 & 4.55 & $-0.63$ \\
Fe~6147& Fe~{\sc ii} & 6147.741 & 3.89 & $-2.72$ \\
Fe~6149& Fe~{\sc ii} & 6149.258 & 3.89 & $-2.72$ \\
Fe~6150& Fe~{\sc ii} & 6150.098 &11.45 & $-3.26$ \\
Fe~6157& Fe~{\sc i} & 6157.725 & 4.08 & $-1.26$ \\
Fe~6705& Fe~{\sc i} & 6705.101 & 4.61 & $-1.50$ \\
Fe~6708& Fe~{\sc ii} & 6708.885 & 10.91 & $-0.52$ \\
\hline
Ni~5682& Ni~{\sc i} & 5682.198 & 4.11 & $-0.47$ \\
\hline
\end{tabular}
\end{center}
\scriptsize
Note. \\
All data are were taken from Kurucz \& Bell's (1995) compilation,
except for those of Li~{\sc i} which were adopted from Smith et al. (1998)
as in Takeda and Kawanomoto (2005).\\
$^{\dagger}$Since this line comprises two close components of stronger 
$\log gf = -0.450$ (at 5688.205~$\rm\AA$) and weaker $\log gf = -1.400$ 
(at 5688.194~$\rm\AA$), we adopted $\log gf = -0.40$ as the sum of these two.
Regarding the Li~{\sc i} lines, we considered only the lines of $^{7}$Li,
since we neglected $^{6}$Li in our analysis.\\
\end{table}

\end{document}